# Significant Phonon Drag Enables High Power Factor in the AlGaN/GaN Two-Dimensional Electron Gas

Ananth Saran Yalamarthy[1], Miguel Muñoz Rojo[2,3], Alexandra Bruefach[4], Derrick Boone[5,6], Karen M. Dowling[2], Peter F. Satterthwaite[2], David Goldhaber-Gordon[6,7], Eric Pop[2,8,9], and Debbie G. Senesky[2,9,10*]

[1]Department of Mechanical Engineering, Stanford University, Stanford, CA 94305, USA.
[2]Department of Electrical Engineering, Stanford University, Stanford, CA 94305, USA.
[3]Department of Thermal and Fluid Engineering, University of Twente, Enschede, 7500 AE, Netherlands.
[4]Department of Materials Science and Engineering, UC Berkeley, CA 94720, USA.
[5]Department of Applied Physics, Stanford University, Stanford, CA 94305, USA.
[6]Stanford Institute for Materials and Energy Sciences, SLAC National Accelerator Laboratory, Menlo Park, CA 94025, USA.
[7]Department of Physics, Stanford University, Stanford, CA 94305, USA.
[8]Department of Materials Science and Engineering, Stanford University, Stanford, CA 94305, USA.
[9]Precourt Institute for Energy, Stanford University, Stanford, CA 94305, USA.
[10]Department of Aeronautics and Astronautics, Stanford University, Stanford, CA 94305, USA.

*Corresponding author: Debbie G. Senesky (dsenesky@stanford.edu)

In typical thermoelectric energy harvesters and sensors, the Seebeck effect is caused by diffusion of electrons or holes in a temperature gradient. However, the Seebeck effect can also have a phonon drag component, due to momentum exchange between charge carriers and lattice phonons, which is more difficult to quantify. Here, we present the first study of phonon drag in the AlGaN/GaN two-dimensional electron gas (2DEG). We find that phonon drag does not contribute significantly to the thermoelectric behavior of devices with ~100 nm GaN thickness, which suppress the phonon mean free path. However, when the thickness is increased to ~1.2 μm, up to 32% (88%) of the Seebeck coefficient at 300 K (50 K) can be attributed to the drag component. In turn, the phonon drag enables state-of-the-art thermoelectric power factor in the thicker GaN film, up to ~40 mW m$^{-1}$ K$^{-2}$ at 50 K. By measuring the thermal conductivity of these AlGaN/GaN films, we show that the magnitude of the phonon drag can increase even when the thermal conductivity decreases. Decoupling of thermal conductivity and Seebeck coefficient could enable important advancements in thermoelectric power conversion with devices based on 2DEGs.

The scattering of electrons and holes by lattice vibrations, known as phonons, often limits the performance of modern transistors and circuits.[1] Yet that same coupling of phonons to charge carriers can also enhance the Seebeck coefficient ($S$), and hence allow increased power generation in thermoelectric (TE) devices.[2–4] Momentum transfer from non-equilibrium phonons to charge carriers, known as phonon drag (PD), produces a Seebeck coefficient ($S_{ph}$) that adds to the Seebeck coefficient from the thermal diffusion of charge carriers ($S_d$). Despite the potential gains in TE efficiency, understanding the contribution of PD to the overall Seebeck coefficient has not received much consideration, largely due to early work which suggested that: (1) $S_{ph}$ is only significant at low temperatures ($T \leq 50$ K), where the TE power conversion efficiency ($zT$) is low;[5] (2) $S_{ph}$ is small relative to $S_d$ for degenerate semiconductors,[6,7] which are the most common TE materials due to their larger $zT$; and (3) an increase in $S_{ph}$ coincides with a corresponding increase the thermal conductivity ($k$),[8–10] and thus has little benefit for power generation, because $zT \propto S^2/k$.

Contrary to these beliefs, recent experiments show that $S_{ph}$ is almost 34% of the total $S$ at room temperature in degenerate, bulk Si (doping of ~$10^{19}$ cm$^{-3}$).[11] Further, recent first-principles calculations show that different ranges of phonon mean free paths (MFPs) contribute to thermal conductivity and PD, respectively. Remarkably, this decoupling means that $k$ could be reduced while preserving $S_{ph}$.[4] This decoupling could be achieved in degenerate two-dimensional electron gases (2DEGs) in semiconductor quantum wells,[12–15] where the 2DEG is confined within a few nanometers of a surface, while the thermal conductivity $k$ is largely determined by phonon scattering within the various layers forming the quantum well.

Previous determinations of $S_{ph}$ in 2DEG systems have relied on measuring the total Seebeck coefficient, theoretically estimating $S_d$, and calculating $S_{ph} = S - S_d$.[16] However, estimating $S_d$ is difficult, requiring precise knowledge of all scattering mechanisms, in addition to the subband energies of the 2D quantum well. In the simple Herring model,[2] $S_{ph} \propto \lambda_{ph}$, where $\lambda_{ph}$ is the MFP of the "representative" phonons contributing to drag. Thus, as shown in recent work on Si,[11] one can separately determine $S$ and $S_{ph}$ by varying the semiconductor dimensions,[17] which controls the distribution of phonon MFPs, and hence $S_{ph}$. As the sample thickness is reduced below a critical value, $S_{ph}$ disappears such that in these samples $S \approx S_d$.[11] $S_{ph}$ in thicker samples can thus be estimated by subtracting the $S_d$ of the smaller samples. Because this method does not rely on a theoretical estimate of $S_d$, it allows for a true extraction of $S_{ph}$, provided that the thickness reduction has minimal effect on the quantum well itself.

In this work, we extend the concept of dimension scaling to extract $S_{ph}$ in the 2DEG that is formed at the surface of a GaN layer (of controlled thickness) capped with a thin, unintentionally doped AlGaN layer. This approach enables the first experimental measurements of $S_{ph}$ in this material system,[4] which is possible up to room temperature given the relatively high Debye temperatures of both GaN and AlN (600 K and 1150 K).[18] In terms of potential applications, this is an appealing heterostructure for use in space environments,[19] where extreme temperature TE power sources[20] are necessary.

Experimental samples were fabricated via metal organic chemical vapor deposition (MOCVD) on a Si (111) wafer (725 μm thick, p-type, doping level of $10^{16}$-$10^{17}$ cm$^{-3}$), as summarized in Supplementary Figure S1. A buffer stack consisting of Al$_x$Ga$_{1-x}$N was grown, followed by a GaN layer whose thickness was chosen to tune the phonon scattering and confinement. Two variants were grown: (i) a "thin" sample with $t_{GaN} \approx 100$ nm and (ii) a "thick" sample with $t_{GaN} \approx 1.2$ μm. The 2DEG was formed by depositing 1 nm/30 nm/3 nm of AlN/Al$_{0.25}$Ga$_{0.75}$N/GaN (cap) on top of the GaN layer, a standard stack for achieving high electron mobility (1500 to 2000 cm$^2$V$^{-1}$s$^{-1}$ at room temperature).[21] The 2DEG forms in GaN at the interface with AlGaN, with a nominal sheet density $n_{2D} \approx 10^{13}$ cm$^{-2}$ and a characteristic quantum well width of ~5 nm.[14] The GaN layer in the two variants is much larger than the quantum well width, which is necessary to ensure that its properties (such as the subband spacing and energies) are not affected. The buffer layers (Al$_x$Ga$_{1-x}$N, $0 \leq x \leq 1$) and the GaN layer are unintentionally doped below $10^{16}$ cm$^{-3}$, ensuring that the measured Seebeck coefficient arises exclusively from the 2DEG.[22]

Extraction of TE properties ($S$ and $k_{GaN}$) is facilitated by inducing a temperature gradient in the plane of the 2DEG. We accomplished this by etching the Si from the backside to create suspended AlGaN/GaN diaphragms, as depicted in Figures 1a and 1b. A 2DEG mesa was then defined by etching off the top AlGaN except in a rectangular strip across which we measured voltage to extract the Seebeck coefficient. After forming a ~47 nm Al$_2$O$_3$ dielectric layer by atomic layer deposition (ALD) to provide electrical isolation from the 2DEG (see Supplementary Note 1), heater electrodes (Pt) were deposited to create an in-plane temperature gradient across the 2DEG mesa. A gate electrode (Au) on top of the Al$_2$O$_3$ (Figures 1a and 1c) enables modulating the charge density in the 2DEG.

Upon applying of a temperature gradient via the Pt heater, a Seebeck voltage is measured across the mesa, which is the sum of thermal diffusion of the 2DEG electrons ($V_d$) and the drag imparted to them by phonons in the GaN layer ($V_{ph}$), as seen in Figure 1c. Using the heater as a thermometer, we extracted the Seebeck coefficient from the voltage across the 2DEG mesa, after accounting for the thermal losses in the $Al_2O_3$ layer and the various interfaces (see Supplementary Note 2). A similar structure with two metal electrodes (heater and sensor) on the suspended AlGaN/GaN diaphragm was used to extract the thermal conductivity of the GaN and the underlying buffer layers. Further details of the measurement process can be found in Supplementary Note 2. The flowchart in Figure 1d details the theoretical calculations and experimental measurements we performed to extract $S_{ph}$. Measurements of the 2DEG sheet density, $n_{2D}$ and mobility, $\mu$ were taken and compared with an analytical model to obtain the energy-dependent scattering times, $\tau(E)$ for electrons in the 2DEG. The obtained $\tau(E)$ is used to calculate the diffusive component of the Seebeck coefficient, $S_d$. The thermal conductivity measurements are used to extract the energy-dependent distribution of phonon scattering lengths in the GaN layer, which is combined with $\tau(E)$ to calculate $S_{ph}$. This modeled $S_{ph}$, along with the calculated $S_d$, can be compared with the experimental values of the Seebeck coefficient for both the thick and thin GaN samples to shed light on the relative contribution of $S_{ph}$.

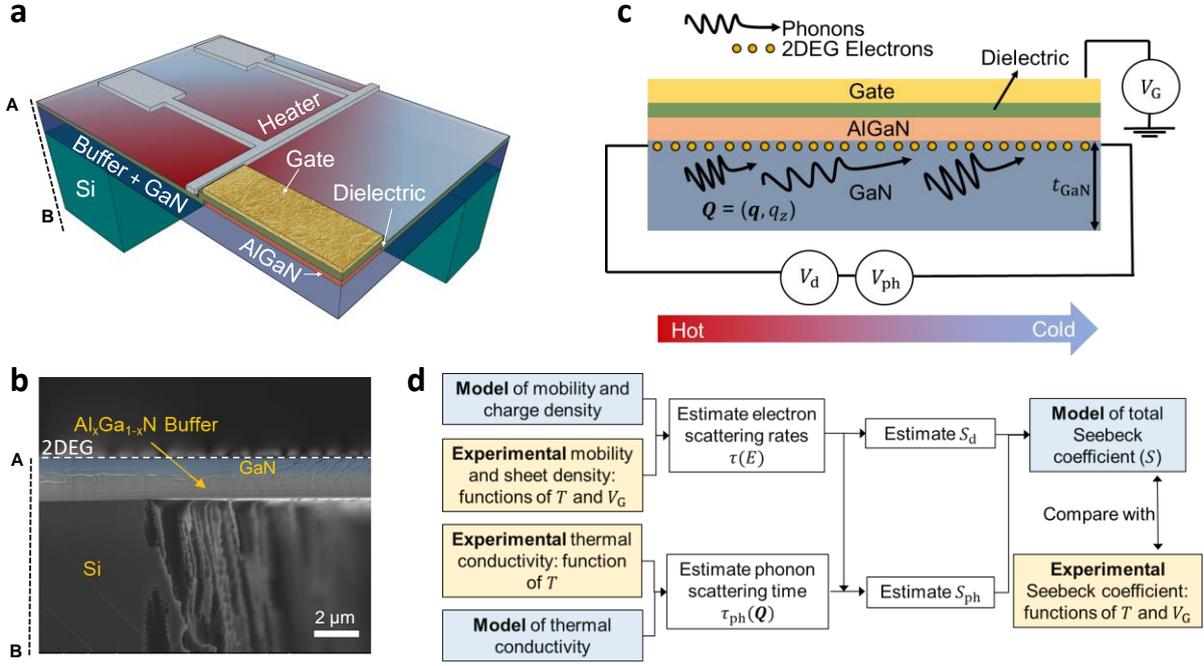

**Figure 1 | Measurement platform to probe 2DEG phonon drag.** (a) Schematic cross-section of suspended device to measure Seebeck coefficient, showing the heater metal, the AlGaN/GaN mesa, and the gate. (b) Cross-sectional SEM image of the suspended region, showing Si, the buffer and the GaN layer. This image is for the thick GaN sample, with $t_{GaN} \approx 1.2$ μm. (c) 2D schematic of the suspended mesa region, showing the drag and diffusive components of the Seebeck voltage. The phonon wave vector is marked by the symbol $\mathbf{Q}$. (d) Flowchart showing the numerical procedure to extract the phonon drag component of the Seebeck coefficient, $S_{ph}$.

We first discuss the measurements of these parameters with the gate grounded. Figure 2a shows measurements of $n_{2D}$ for the thick and thin GaN sample, extracted via Hall effect and van der Pauw measurements. The inset shows a schematic band diagram of the AlGaN/GaN quantum well, with the 2DEG depicted as the triangular region at the interface below the Fermi level ($E_F$). The thickness of the quantum well, $t_{2D}$, is defined as the distance from the AlGaN/GaN interface to the intersection of $E_F$ and the GaN conduction band. In both samples, we obtain sheet density $n_{2D}$ roughly independent of temperature from 50 K to 300 K, consistent with the weak temperature dependence of the piezoelectric constants of both AlN and GaN.[23] The thin and thick GaN samples have a similar $n_{2D} \approx 10^{13}$ cm$^{-2}$,[14] verified using a commercially available Schrödinger-Poisson solver[24] as seen in Figure 2a. We also obtain $t_{2D} \approx 6.1$ nm and $t_{2D} \approx 4.4$ nm for the thick and thin GaN sample from the solver. For simplicity, in the models for TE transport properties we set $n_{2D} = 10^{13}$ cm$^{-2}$ for both samples.

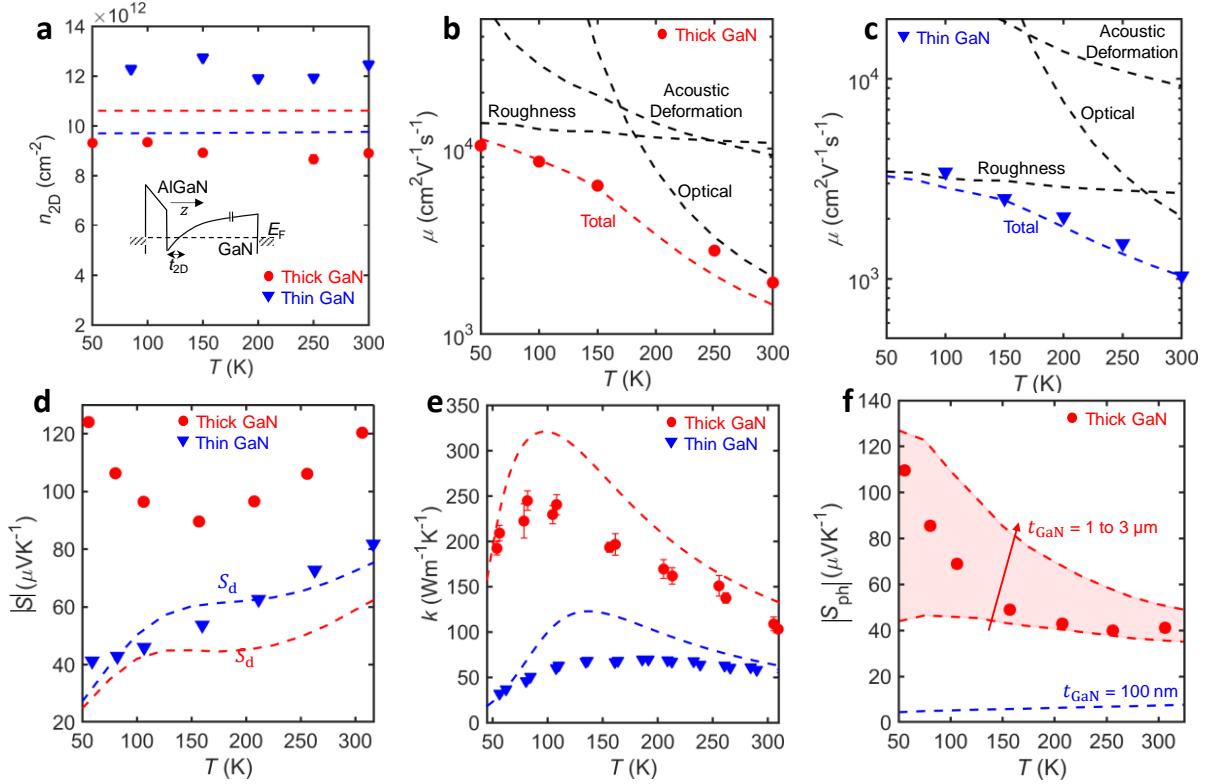

**Figure 2 | Thermoelectric property measurements.** (a) Temperature dependent sheet density ($n_{2D}$) of the thick and thin GaN sample. The experimental markers (blue triangles and red circles) are obtained from Hall-effect and van der Pauw measurements, while the dashed lines show the simulated values obtained from a commercial solver. The inset shows a schematic of the AlGaN/GaN quantum well, with the Fermi level and the characteristic thickness of well, $t_{2D}$, marked. (b,c) Mobility for the thick and thin GaN sample, with the dashed lines showing the simulated components, and the markers from Hall and van der Pauw measurements. (d) Measured Seebeck coefficient. The dashed lines show the calculated diffusive components, which are similar for the thick and thin GaN samples. (e) Measured (markers) and calculated (dashed lines) thermal conductivities for the thick and thin GaN samples. (f) Simulated values of the phonon drag component of the Seebeck coefficient obtained by sweeping the effective thickness of the GaN layer. The red markers show the estimated drag component for the thick GaN sample extracted from the experimental data. A clear suppression of phonon drag is observed for smaller GaN layer thickness.

Using the expression for the 2D density of states, assuming that all the sheet density is from a single subband, $g_{2D} = \frac{m^*}{\pi\hbar^2}$, we obtain $E_F - E_1 \approx 110$ meV, where $E_1$ denotes the energy at the bottom of the first subband. Here, $m^*$ is the electron effective mass in GaN (Table S1). This is consistent with the energies obtained from the solver (Supporting Note 3), and indicates that only the bottom subband contributes significantly to charge density. For the rest of this work, only this bottom subband is considered in the calculation of the Seebeck coefficient.[25]

Next, we turn to measurements of the 2DEG mobility obtained via Hall-effect, plotted with symbols in Figure 2b and Figure 2c for the thick and thin GaN samples, respectively. The dashed lines show the calculated contributions to the mobility from scattering mechanisms that are dominant in AlGaN/GaN 2DEGs.[26] Other scattering mechanisms (e.g. dislocation, ionized impurity and piezoelectric scattering) are neglected. Rigorous justification of this approximation is found in Supplementary Note 2. For both thick and thin GaN, polar optical phonon (POP) scattering is the dominant scattering mechanism at room temperature, due to the large optical phonon energy ($\hbar\omega_{OP} = 91.2$ meV),[27] and the polar nature[28] of the GaN wurtzite crystal. Though the optical phonon population decreases exponentially at lower temperatures, electrons in the lower subband still scatter against the AlGaN/GaN interface roughness. To estimate this component, we set the root-mean-square (RMS) roughness height, $\Delta = 1$ and 2 nm for the thick and thin GaN sample, respectively (atomic force microscopy of the sample surface can be found in Supplementary Figure S4). The good agreement between the model and experimental data allows us to extract the energy-dependent scattering time, $\tau(E)$ for electrons in the bottom subband of the 2DEG.

From this, we can calculate the diffusive component of the Seebeck coefficient for the bottom subband[29]

$$S_d = \frac{-1}{eT} \frac{\int E \frac{\partial f_0(E)}{\partial E}(E - E_F - E_1)\tau(E)dE}{\int E \frac{\partial f_0(E)}{\partial E}\tau(E)dE}, \tag{1}$$

where $f_0(E)$ is the equilibrium Fermi function, and $e$ is the magnitude of the electronic charge. These are plotted against the experimental data for the magnitude of the Seebeck coefficient (the actual sign is negative) in Figure 2d. The theoretical curves deviate slightly from a linear dependence on temperature, typical for a degenerate semiconductor.[25] This deviation is due to POP scattering, which forbids electrons with energies smaller than $\hbar\omega_{OP}$ from emitting optical phonons.[29] The slight difference in the calculated values of $S_d$ for the thick and thin GaN sample is found to arise from the difference in the roughness scattering component of $\tau(E)$. We observe that the Seebeck coefficient for the thin GaN sample agrees well with the calculated $S_\text{d}$, however this model cannot describe the thick GaN sample (Figure 2d). In addition, the magnitude of the Seebeck coefficient in the thick GaN sample exhibits a prominent upturn at low temperatures, hinting at PD.[16]

In our device, three-dimensional (3D) phonons, represented by the wave vector $\boldsymbol{Q} = (\boldsymbol{q}, q_z)$, which represent the in-plane (of the 2DEG) and out-of-plane component, scatter with 2D electrons in the bottom subband, giving rise to $S_\text{ph}$. To calculate this drag, we follow the approach introduced by Cantrell and Butcher[3] and later modified by Smith.[30,31] We explicitly include the dependence of phonon scattering time ($\tau_\text{ph}$) on the phonon wave vector

$$S_\text{ph} = -\frac{(2m^*)^{\frac{3}{2}}v_\text{av}^2}{4(2\pi)^3 k_\text{B} T^2 n_\text{2D} e\rho} \int_0^\infty dq \int_{-\infty}^\infty dq_z \frac{\Xi^2(\boldsymbol{Q}) q^2 Q^2 |I(q_z)|^2 G(\boldsymbol{Q}) \tau_\text{ph}(\boldsymbol{Q})}{S^2(q,T) \sinh^2\left(\frac{\hbar\omega_Q}{2k_\text{B}T}\right)}. \quad (2)$$

In Equation 2, $v_\text{av}$ is the average phonon velocity over the different modes, $k_\text{B}$ is the Boltzmann constant, and $\rho$ is the mass density of GaN. Values of the parameters used for our calculations are in Supplementary Table S1. The phonon frequency, $\omega_Q$ is approximated as $v_\text{av}\sqrt{q^2 + q_z^2}$ assuming a 3D isotropic linear dispersion. The term $I(q_z) = \int \psi(z)^2 e^{iq_z z}$ describes the electron-phonon momentum conservation in the $z$ direction, where $\psi(z)$ is the wave function of the electrons in the bottom subband. $\Xi(\boldsymbol{Q})$ represents the strength of the electron-phonon coupling. The terms $S(q,T)$ and $G(\boldsymbol{Q})$ represent a screening function for the electrons and an energy integral, respectively (the detailed explanation of these terms is discussed in Supplementary Note 4). Of particular interest to this work is $\tau_\text{ph}(\boldsymbol{Q})$, representing the phonon relaxation time. This term describes the scaling dependence of $S_\text{ph}$ on sample thickness, because $\tau_\text{ph}(\boldsymbol{Q}) \propto t_\text{GaN}$ due to boundary scattering.

To calculate $\tau_\text{ph}(\boldsymbol{Q})$ accurately, we measured the thermal conductivity, $k$, of the suspended diaphragms, presented in Figure 2e. Because our suspended film is a composite consisting of an AlN layer, Al$_x$Ga$_{1-x}$N transition layers and a GaN layer, the overall thermal conductivity must be estimated from an average of the thermal conductivities, weighted by the thicknesses of individual layers. For each layer, we used a Boltzmann Transport Equation (BTE) model to quantify its thermal conductivity. The dashed lines in Figure 2e show the modeled $k$ for the entire stack, taking into account phonon-phonon, dislocation, alloy and boundary scattering using standard values of the elastic moduli for AlN and GaN (details in Supplementary Note 5). This use of standard values of the elastic moduli, alloy scattering, and dislocation scattering terms, which are challenging to obtain experimentally,[32,33] could explain the disagreement between the model and the data. Yet, this model will suffice to explain the observed trends in the PD behavior. Assuming that only the phonons in the GaN layer contribute to drag, the modelled $\tau_{ph}$ for this layer is combined with Equation 2 to calculate $S_{ph}$.

The modeled $|S_{ph}|$ is plotted in Figure 2f for a range of effective GaN thicknesses ($t_{GaN}$) values. The magnitude of $S_{ph}$ (actually negative in sign) for the thin GaN is between 4 and 8 µVK$^{-1}$ across all $T$, significantly less than the measured 40 to 80 µVK$^{-1}$ (Figure 2d), supporting the conclusion that $S \approx S_d$. The near temperature-independence of the modeled $S_{ph}$ is due to $k_{GaN}$ being limited by boundary scattering across the entire temperature range. $S_{ph}$ in the thick GaN film was estimated by subtracting a linear fit (including the origin) of the measured Seebeck coefficient in the thin GaN sample from the measured Seebeck coefficient of the thick GaN sample. We have used a linear fit including the origin of the thin GaN Seebeck coefficient to avoid overestimating the diffusive component of the Seebeck coefficient. This is because the measured Seebeck coefficient values of the thin GaN sample still includes a small PD component, which is visible as a slight flattening at the lower temperatures (blue triangles in Figure 2d).

The estimate of $S_{ph}$ for the thick GaN sample after subtraction from the linear fit is plotted in Figure 2f (red markers). The shaded region shows the calculated $S_{ph}$ for various $t_{GaN}$ from 1 to 3 µm using Equation 2. We have swept the GaN thickness in the model because it under-predicts $S_{ph}$ if we use the actual thickness (1.2 µm). This need to introduce an effective parameter may arise from the simple model for the thermal conductivity and PD used here, and the difficulty in determining the 2DEG quantum well thickness experimentally. The model exhibits the correct trend within the swept thickness range. The experimental $S_{ph}$ data (red circles in Figure 2f) show that ~32% of the total $S$ at room temperature can be attributed to drag, increasing to almost 88% of $S$ at 50 K for the thick GaN sample. The inverse temperature dependence of $S_{ph}$ is reflective of phonon-phonon scattering, from which the phonon MFP scales as $T^{-1}$. The measurements of the Seebeck coefficient and the thermal conductivity for the thick GaN sample below ~90 K (red circles) in Figures 2e and 2f also suggest that the PD continues to increase even when the thermal conductivity starts decreasing. This provides experimental evidence that these two parameters can be decoupled to increase *zT*, in agreement with previous theoretical work.[34,35]

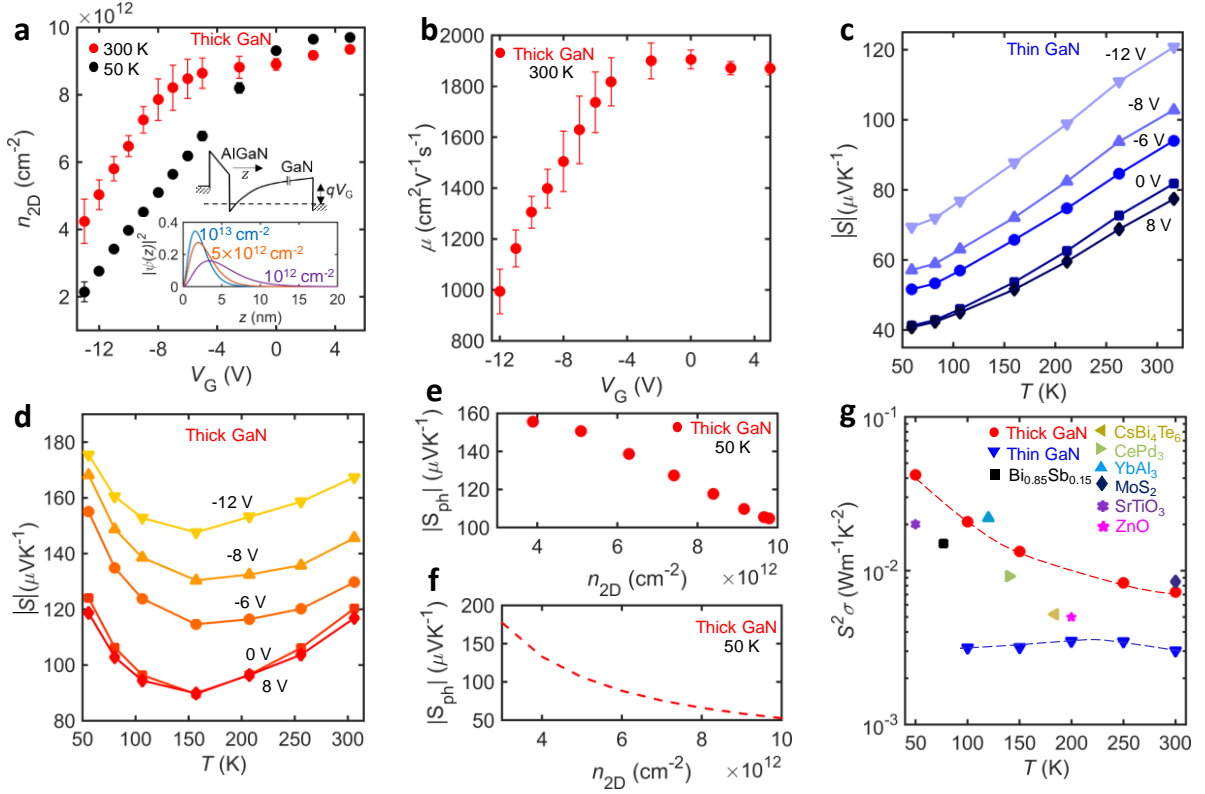

**Figure 3 | Measurements with a gate bias.** (a) Modulation of the sheet density in the 2DEG ($n_{2D}$) with applied gate bias at 300 K and 50 K for the thick GaN sample. The markers are obtained from Hall-effect measurements. The inset shows the simulated wave function in the bottom subband of the 2DEG for three different sheet densities. The coordinate $z=0$ corresponds to the AlGaN/GaN interface, as seen in the band diagram (black lines). Positive $z$ represents the GaN layer. (b) Experimental measurements of field-effect mobility at 300 K. (c,d) Gated Seebeck coefficient measurements for the thin and thick GaN sample. The solid lines are a guide for the eye, while the makers are the experimental measurements. (e) Estimated drag component for the thick GaN sample from the experimental data, at 50 K. (f) Simulated phonon drag component for the thick GaN sample (using $t_{GaN} = 1.2$ μm) for 2DEG sheet densities that correspond to our applied voltage range, at 50 K. (g) Estimated temperature-dependent power factors for the thin and the thick GaN samples from 50 to 300 K from the experimental data, with the gate grounded. The dashed red and blue lines are guidelines for the eye. We have also included power factor data from other material systems for comparison with the thick GaN 2DEG.

The application of a gate voltage, $V_G$, can tune the TE power factor ($S^2\sigma$) without changing $k$, which can further optimize $zT$.[36,37] While the effect of $V_G$ on $S_d$ is well known, only a few studies have attempted to quantify its effect on drag.[16,38,39] In particular, application of $V_G$ tunes the quantum well width and 2DEG charge density ($n_{2D}$), simultaneously. $S_{ph}$ is inversely proportional to $n_{2D}$ giving it a strong dependency on this parameter, as seen in Equation 2. Quantum well width affects $S_{ph}$ through $I(q_z)$ which is strongly dependent on the wave function $\psi(z)$. A more tightly confined wave function in real space (which corresponds to larger $n_{2D}$) is broader in Fourier space, increasing $I(q_z)$. These two effects compete against each other, resulting in a complex gate voltage dependency.

Hall-effect measurements of the 2DEG sheet density as a function of gate voltage are presented in Figure 3a. The data at 300 K shows a depletion of the 2DEG sheet density by up to a factor of ~3x from its ungated value as $V_G$ is lowered to -12 V. The gating is similar at lower temperatures (data for the thick GaN sample at 50 K are plotted with black circles in Figure 3a) and for the thin GaN sample. The inset of Figure 3a shows how depletion widens the quantum well at the AlGaN/GaN interface. Depletion also reduces the 2DEG mobility as seen in Figure 3b, similar to former work.[40,41] To study the effect of gating on $S_{ph}$, we need to first estimate $S_d$ as a function of gate voltage. This can be done by studying the effect of $V_G$ on the thin GaN sample, presented in Figure 3c. For a degenerate 2D quantum well, we can roughly approximate the magnitude diffusive Seebeck coefficient as $S_d \propto T/(E_F - E_1)$.[25] Since $n_{2D} \propto (E_F - E_1)$, the magnitude of the diffusive Seebeck coefficient should increase as negative $V_G$ depletes the 2DEG, and decrease linearly with $T$. Both features are visible in Figure 3c.

Figure 3d shows the effect of $V_G$ on |S| in the thick GaN sample, where the upturn below ~150 K is apparent even after depletion. As in Figure 2f, we subtracted a linear fit of the thin GaN Seebeck coefficients (in Figure 3c) from the values for the thick GaN to estimate $S_{ph}$ for different $V_G$. Because we know the relation between $n_{2D}$ and $V_G$ (Figure 3a), we can thus estimate $S_{ph}$ as a function of $n_{2D}$. We have plotted the gate-voltage dependence of $S_{ph}$ at a fixed temperature of 50 K, for different $n_{2D}$ values in Figure 3e. It is seen that $|S_{ph}|$ increases by a factor of ~1.5x as $n_{2D}$ decreases from $10^{13}$ cm$^{-2}$ to $3\times10^{12}$ cm$^{-2}$. To confirm the trend of these values, we also simulated $S_{ph}$ over this $n_{2D}$ range using Equation 2 (with the actual GaN thickness of 1.2 $\mu$m), taking into account the shape of the quantum well. These simulations are plotted in Figure 3f at a temperature of 50 K, for ease of comparison to the data in Figure 3e. The simulated data shows the same trend (i.e., $|S_{ph}|$ increasing as $n_{2D}$ decreases), but the increase is much larger (~3x). Although the reason for the mismatch needs further study, these trends of $S_{ph}$ vs. $V_G$ suggest that the Seebeck coefficient behavior in the thick GaN sample is indeed due to PD. Further, they show that depleting the AlGaN/GaN 2DEG increases the magnitudes of both the diffusive and drag components of the Seebeck coefficient.

Finally, it is worthwhile to examine the TE power factor ($S^2\sigma$) of the 2DEG in both the thick and the thin GaN sample. These values are plotted in Figure 3g, where the gate is grounded. In order to calculate the conductivity of the 2DEG, $\sigma$, we use the mobility values in Figure 2b and Figure 2c, along with an estimate for the average volumetric charge density, $n_v = n_{2D}/t_{2D}$.[41] The $n_{2D}$ values are taken from the experimental values in Figure 2a. While the power factor for the thin GaN sample is quite insensitive to temperature, the value for the thick GaN sample shows a pronounced enhancement at low temperatures, as seen in Figure 3g, reaching ~40 mW m$^{-1}$ K$^{-2}$ at 50 K. This high power factor, which originates from the upturn of the Seebeck coefficient at low temperatures via PD, is state-of-the-art when compared with other TE materials also plotted in Figure 3g ($Bi_{0.85}Sb_{0.15}$,[42] $CsBi_4Te_6$,[43] $CePd_3$,[44] $YbAl_3$,[45] $MoS_2$[46]). We have also plotted the power factors for other 2DEG systems where measurements are available, such as gated ZnO[37] and gated $SrTiO_3$[38] for comparison in Figure 3g. The enhancement of the Seebeck coefficient in our thick GaN sample is in contrast with typical TE materials, where the power factor scales directly with temperature because the Seebeck coefficient is diffusive.[43] The high power factor values in the thick GaN sample, although only for a single 2DEG, are promising for planar applications such as Peltier coolers. Further, they could make promising low-temperature energy harvesting elements when structured as a superlattice.[47]

In conclusion, we have experimentally shown that PD can be a significant portion of the total Seebeck coefficient in a 2DEG, even at room temperature. By using thickness as a "knob" to control sample dimensions, we show that $S_{ph}$ is suppressed in the AlGaN/GaN 2DEG at a film thickness of ~100 nm. From a TE power conversion perspective, we shed light on two important phenomena: First, the magnitude of the PD can increase even when the thermal conductivity is decreasing, which means that these could be tuned separately. Second, depleting a 2DEG can lead to an increase in both the PD and diffusive contributions of the Seebeck coefficient. These findings enable a better understanding of the PD effect, and can lead to advancements in TE power conversion across a wide range of temperatures.

**Supporting Information**

The Supporting Information is available free of charge on the ACS Publications website. Fabrication Process. Test Setup and Measurement Notes. Mobility Model. Phonon Drag Model. Thermal Conductivity Model. Simulation Codes.

**Acknowledgements**

This work was supported in part by the National Science Foundation (NSF) Engineering Research Center for Power Optimization of Electro Thermal Systems (POETS) under Grant

EEC-1449548, and by the NSF DMREF grant 1534279. The MOCVD experiments were conducted at the MOCVD Lab of the Stanford Nanofabrication Facility (SNF), which is partly supported by the NSF as part of the National Nanotechnology Coordinated Infrastructure (NNCI) under award ECCS-1542152. Hall measurements were supported by the U.S. Department of Energy, Office of Science, Basic Energy Sciences, Materials Sciences and Engineering Division, under Contract DE-AC02-76SF00515. D.B.'s participation in this research was facilitated in part by a National Physical Science Consortium Fellowship and by stipend support from the National Institute of Standards and Technology.

# Supporting Information

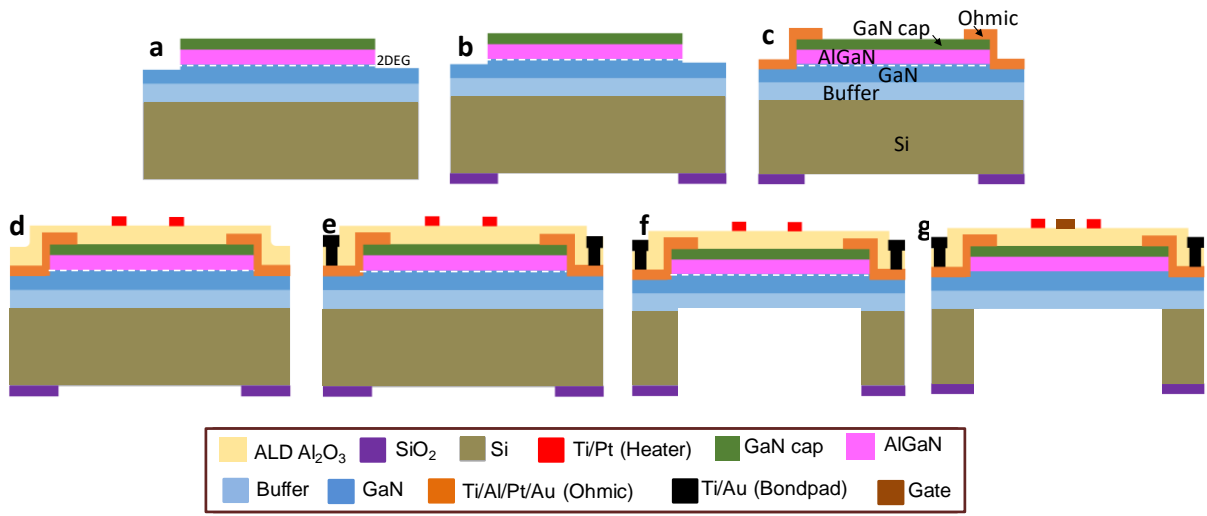

**Figure S1 | Outline of fabrication process.** The panels show the eight mask process to fabricate the suspended thermal conductivity and Seebeck coefficient measurement platforms. This fabrication process is similar to our earlier work.[1]

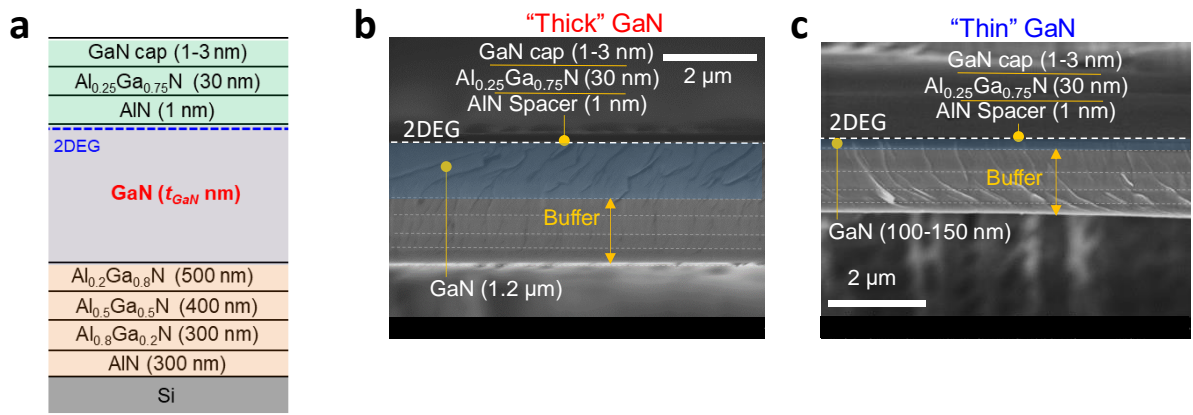

**Figure S2| Cross-sectional images.** (a) Schematic of grown AlGaN/GaN heterostructure, along with the different buffer layers. (b,c) SEM image of the suspended portion of the thick and thin GaN sample.

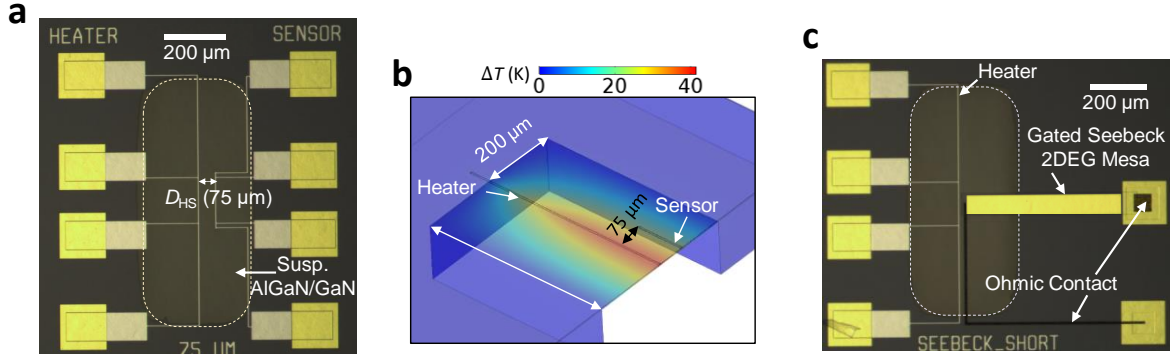

**Figure S3| Test Structures for thermal conductivity and Seebeck coefficient measurement.** (a) Microscope image of the suspended thermal conductivity measurement structure. (b) Half-symmetric finite-element simulation of the thermal conductivity measurement structure, showing sample temperature profile when current is applied through the heater.[1] (c) Microscope image of the suspended Seebeck coefficient measurement structure.

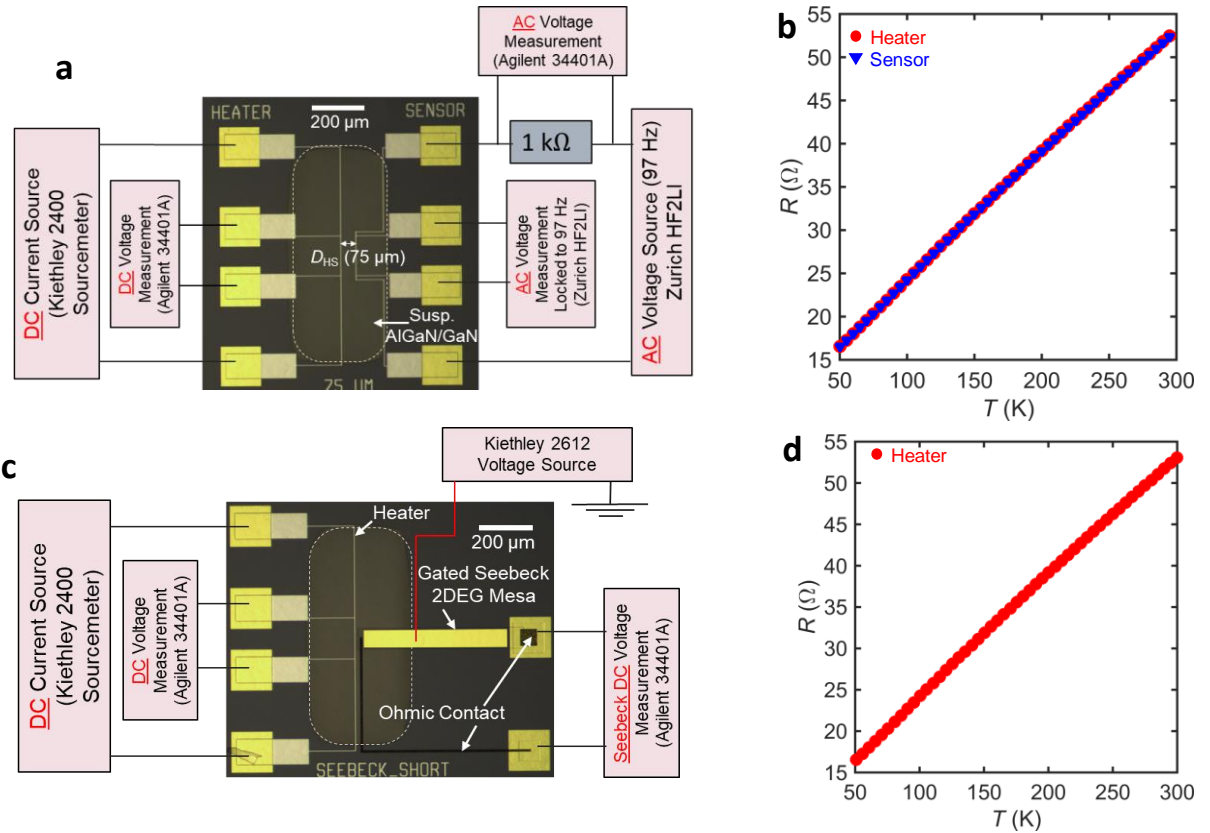

**Figure S4| Measurement Setups for thermal conductivity and Seebeck coefficient measurement.** (a) Schematic of measurement setup to determine the thermal conductivity of the suspended heterostructure layers. We measured the resistance of the heater electrode using a DC multimeter and a DC current source, with a current of 100 µA. For measuring the sensor resistance, we used a lock-in amplifier with a frequency of 97 Hz to minimize self-heating effects. (b) Temperature-resistance calibration for the heater and sensors lines for thermal conductivity measurement. (c) Schematic of measurement setup to determine the Seebeck coefficient of the gated 2DEG mesa. (d) Temperature-resistance calibration of the heater line for Seebeck coefficient measurement.

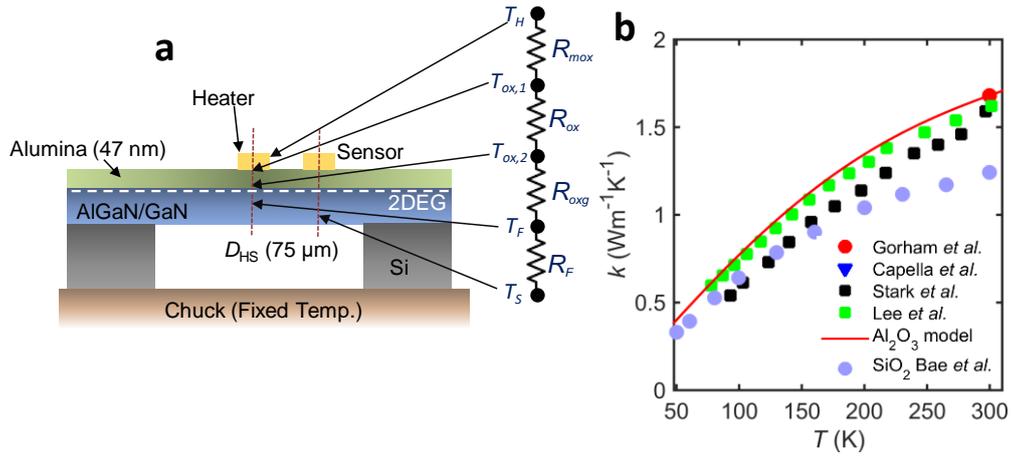

**Figure S5| Details of thermal conductivity measurement.** (a) Cross-section schematic of the thermal conductivity measurement platform, showing the different pathways for heat sinking. (b) Model for the thermal conductivity of alumina, extracted from experimental data in the literature.

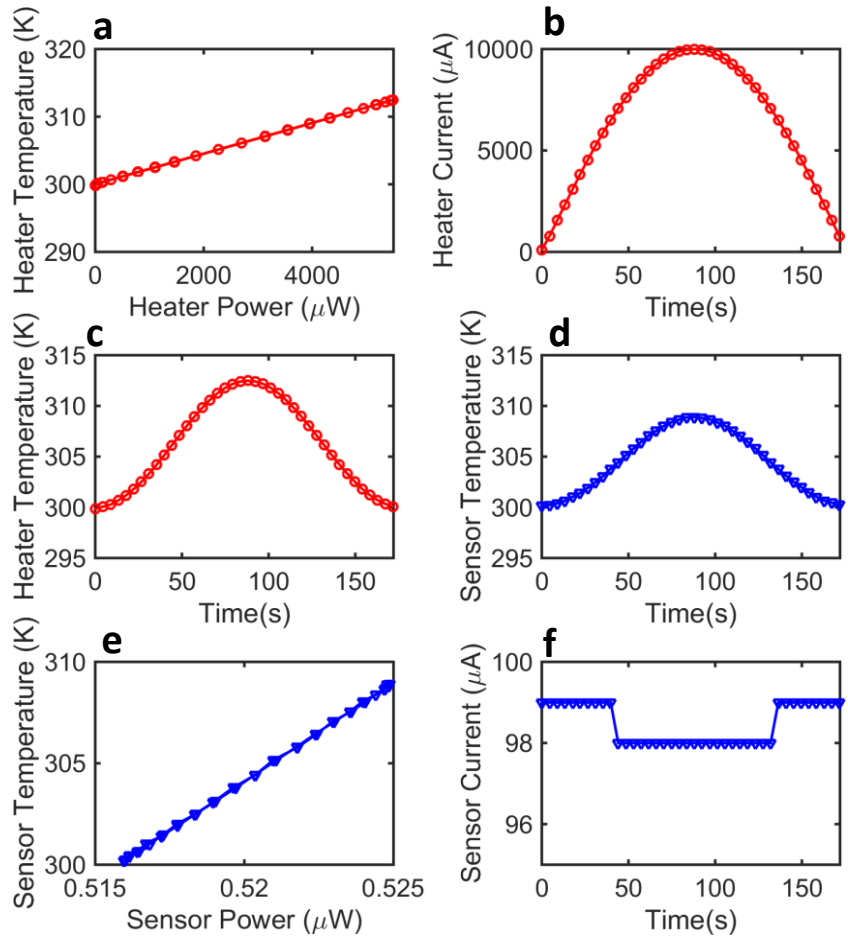

**Figure S6| Thermal conductivity measurement.** Panels (a)-(c) are for the heater line, while panels (d)-(f) are for the sensor line. These panels in this specific example for the thick GaN sample, with the substrate held at 300 K.

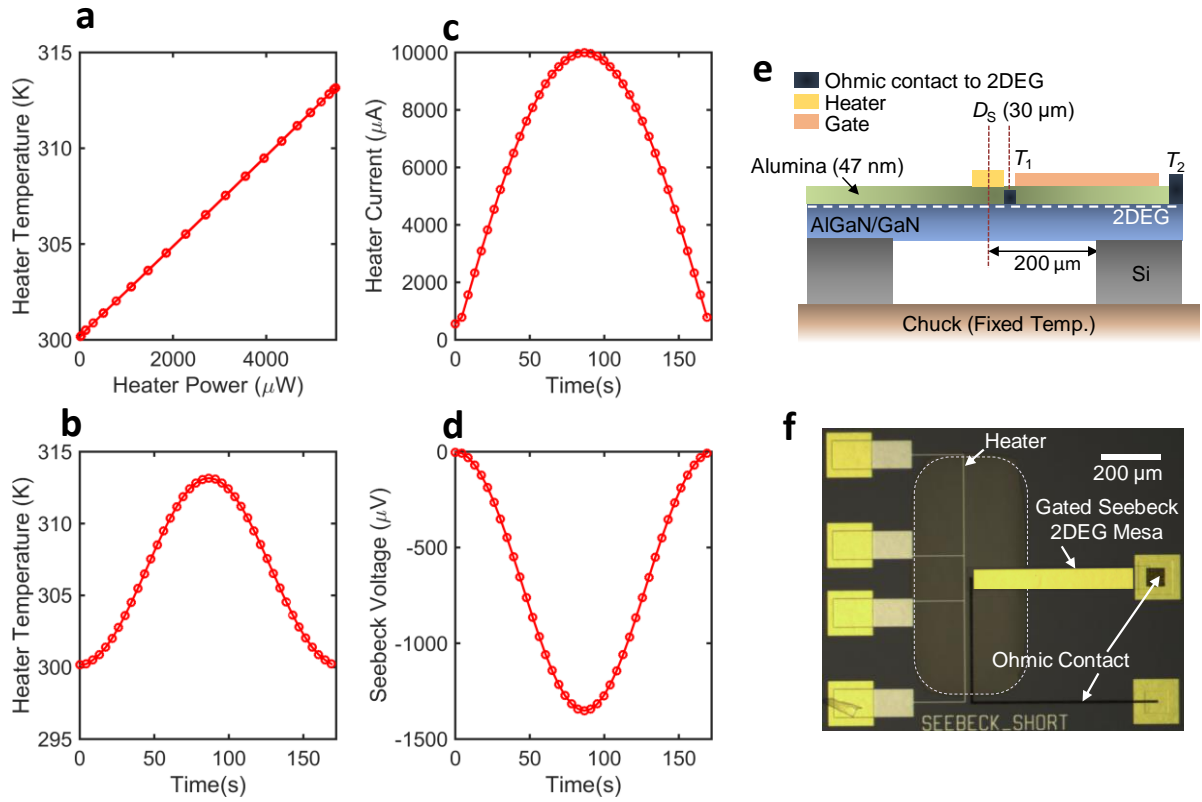

**Figure S7| Seebeck coefficient measurement.** Panels (a)-(c) are for the heater line, while (d) shows the Seebeck voltage measured in the 2DEG mesa. These panels in this example are for the thick GaN sample, with the substrate held at 300 K, and the gate grounded. (e,f) Cross-section and top view showing the different electrodes for Seebeck coefficient measurement.

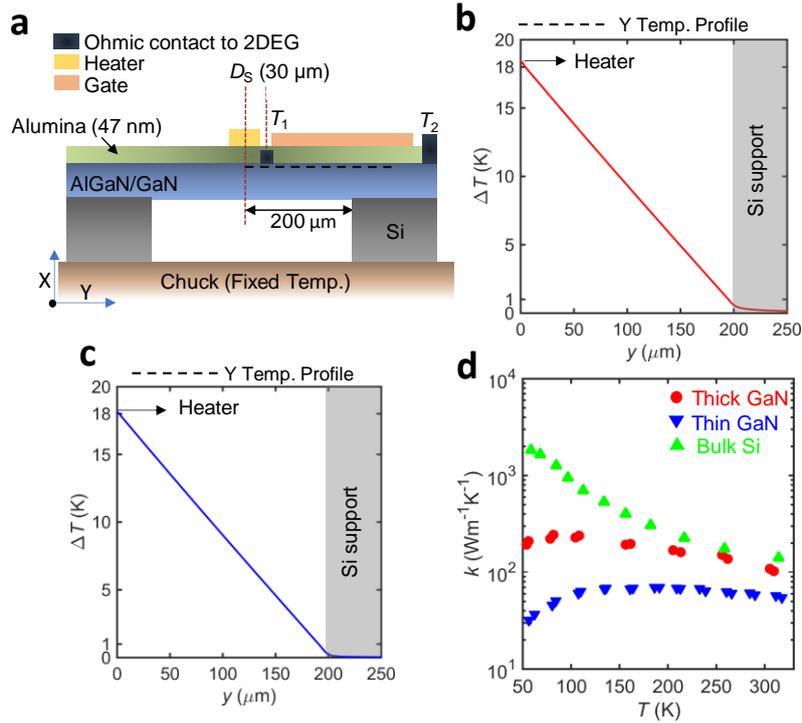

**Figure S8| Details of Seebeck coefficient measurement.** (a) Cross-section schematic of the Seebeck coefficient measurement structure. (b) Simulated temperature drop from the center of the heater electrode to the Si supported region in the thick GaN sample. with the base held at 300 K. The simulation temperature profile is extracted along the black dashed line shown in panel (a), just below the alumina layer in the AlGaN/GaN heterostructure. The simulated current in the heater line is 16 mA. (c) Simulated temperature drop from the center of the heater electrode to the Si supported region in the thin GaN sample. with the base held at 300 K. The simulated current in the heater line is 9 mA. (d) Extracted thermal conductivity of the thick and thin GaN samples, compared with thermal conductivity data for bulk Si (p-type, boron doped, ~$4\times10^{16}$ cm$^{-3}$) from the literature.[2]

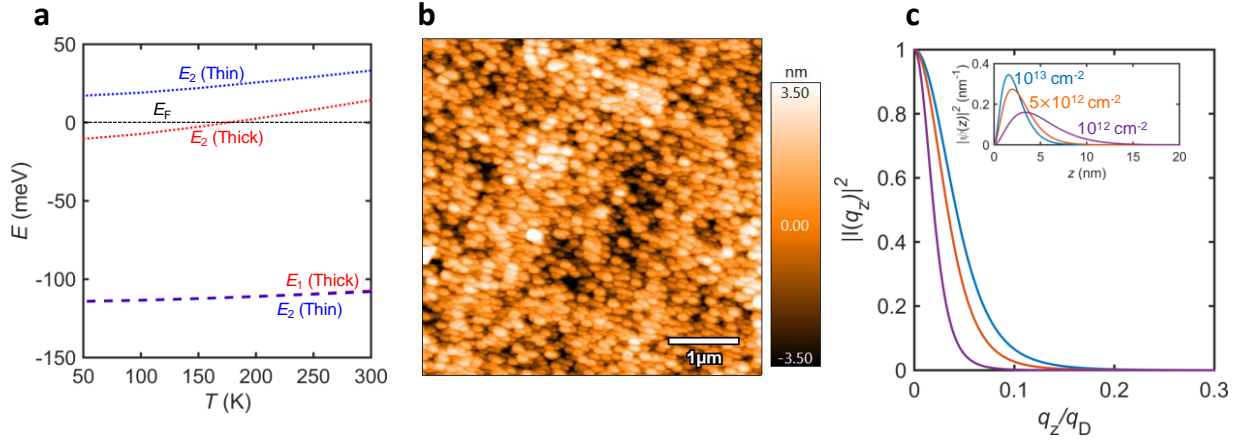

**Figure S9|** (a) Simulated energies of the bottom 2 subbands in the AlGaN/GaN quantum well for the thick and thin GaN samples. (b) AFM image of the surface of the thick GaN sample (with the alumina layer on top). The RMS roughness is estimated to be ~1.4 nm. (c) Electron-phonon momentum conservation in the out-of-plane direction for 2DEG sheet densities varying from 1-10 × $10^{12}$ cm$^{-2}$. The corresponding shape of the wave function, $\psi(z)$, for the bottom subband at the 2DEG at the AlGaN/GaN interface is shown in the inset.

**Table S1| Definitions of selected parameters.**

| Parameter | Symbol (units) | Value | Reference |
|---|---|---|---|
| Effective electron mass | $m^*$ | $0.22m_e$ | Gurusinghe et al.[3] |
| GaN dielectric constant | $\epsilon$ (Fm$^{-1}$) | $10.4\epsilon_0$ | Gurusinghe et al.[3] |
| GaN sheet density | $n_{2D}$ (cm$^{-2}$) | ~1×$10^{13}$ cm$^{-2}$ | Our measurements |
| GaN deformation potential | $D$ (eV) | 8.5 | Sztein et al.[4] |
| Optical phonon energy | $\hbar\omega_{OP}$ (meV) | 91.2 | Sztein et al.[4] |
| Density of GaN, AlN | $\rho$ (kgm$^{-3}$) | 6150, 3266 | Sztein et al.[4] |
| Grüneisen parameter | $\gamma_G$ | 0.5 | Sztein et al.[4] |
| Atomic mass of GaN, AlN | $M$ (amu) | 83.7, 40.99 | Sztein et al.[4] |
| Average phonon velocity in GaN, AlN | $v_{av}$ (ms$^{-1}$) | 5070.5, 7183.5 | Sztein et al.[4] |
| GaN, AlN Debye temperature | $\theta_D$ (K) | 600, 1150 | Sztein et al.[4] |
| GaN Umklapp scattering constants | $P$ (eV), $C_U$ (K) | 1.1375 eV, 132 K | Cho. et al.[5] |
| AlN Umklapp scattering constants | $P$ (eV), $C_U$ (K) | 2.0625 eV, 382 K | Slack et al.[6] |

**Supplementary Note 1:** Fabrication Process

Figure S1 shows the eight-mask process to fabricate the fully-suspended AlGaN/GaN platform for thermal measurements. A schematic of the heterostructure showing the different buffer layers and the silicon substrate is illustrated in Figure S2a. The AlGaN/GaN/buffer heterostructure for the thin and bulk GaN samples was grown using an in-house metal organic chemical vapor deposition (MOCVD) chamber on a 725 μm thick Si(111) substrate (p-type, doping level of $10^{16}$-$10^{17}$ cm$^{-3}$). In order to define the 2DEG mesa, we etched the AlGaN/GaN layers to a depth of ~50 nm using an inductive coupled plasma technique with $BCl_3$/$Cl_2$ gases as shown in Figure S1a. This was followed by the deposition of ~4 μm PECVD oxide on the backside and selectively patterned to define the Si removal region, as depicted in Figure S1b. The Ohmic contacts to the 2DEG were patterned by depositing Ti/Al/Pt/Au (20/100/40/80 nm) followed by a rapid thermal anneal (RTA) in $N_2$ ambient at 850°C for 35 seconds (Figure S1c). Next, we deposited ~47 nm of atomic-layer deposited (ALD) $Al_2O_3$ followed by patterning Ti/Pt (10/100 nm) heater and sensor metal lines, as shown in Figure S1d. To deposit Ti/Au (20/200 nm) bondpad metal, we opened vias in the ALD film using a 20:1 buffered oxide etch for ~2 min (Figure S1e). The gate metal Ti/Au (20/200 nm) was deposited after the bondpad metal, as shown in Figure S1f. To release the AlGaN/GaN/buffer heterostructure, Si was finally etched from the backside using a deep reactive ion etching (DRIE) technique, stopping at the buffer/Si interface. SEM images of the suspended portion of the thick and thin GaN sample are shown in Figure S2b and Figure S2c, respectively. After suspension, the total thickness of the heterostructure layers was obtained as ~2.85 μm for the thick GaN heterostructure and ~1.695 μm for the thin GaN heterostructure and heterostructure from the SEM cross-section images, shown in Figure S2b and Figure S2b.

**Supplementary Note 2:** Test Setup and Measurement Notes

To obtain the thermal conductivity and Seebeck coefficient of the AlGaN/GaN 2DEG, we used a measurement procedure that is similar to our earlier work.[1] All our experiments (from 50 to 300 K) were done in vacuum using a temperature controlled cryostat. To obtain the gate and temperature dependent sheet density, we performed Hall effect and Van der Pauw measurements in a vacuum cryostat. The use of vacuum ensures that any errors in the extraction of the thermal conductivity and Seebeck coefficient due to thermal convection effects are eliminated. In what follows, we briefly outline our method to measure the Seebeck coefficient and the thermal conductivity of our heterostructure layers, followed by a detailed description of the nuances of our measurement scheme.

Figure 2a and Figure 2c are microscope images of our two fully-suspended heterostructure platforms for the measurement of in-plane thermal conductivity of the heterostructure stack and Seebeck coefficient of the 2DEG. As can be seen in Figure 2a, two parallel, ~5 μm wide Ti/Pt metal lines separated by 75 μm are used as heater and sensor thermometers, patterned on a ~47 nm thick amorphous $Al_2O_3$ layer that provides electrical isolation from the heterostructure. For Seebeck coefficient measurement, only a heater thermometer is patterned adjacent to a gated 2DEG mesa with Ohmic contacts extending to the substrate, as illustrated in Figure 2c.

Measurement of the in-plane thermal conductivity is conducted as follows. The sample is attached to a temperature controlled chuck inside a cryostat via a vacuum-compatible thermal grease (Apiezon Inc.) with vacuum as the ambient. We pass a range of DC currents through the heater metal line to induce a temperature gradient in the heterostructure and simultaneously measure the electrical resistance of the metal electrodes. Typical current values are chosen to induce a maximum $\Delta T \sim 20$ K referenced to the substrate temperature, which varies from 50 K to 300 K. The placement of the sensor electrode was carefully designed to allow for a one-dimensional (1-D) in-plane heat transfer approximation in the diaphragm. Figure 2b shows a simulated example of the temperature profile for a half-symmetry region in the thermal conductivity measurement structure, where it can be seen that the heat transfer is 1-D in the center of the membrane. The electrical resistance of the electrodes was calibrated over the entire temperature range (50 to 300 K) using sufficiently low currents to avoid self-heating. The calibration allows us to convert the electrical resistance into corresponding temperature values using the measured temperature-resistance data. From the collected temperature data, we can infer the in-plane thermal conductivity of the heterostructure given the heater power ($P_H$) after accounting for losses due to heat spreading into the $Al_2O_3$ insulation below the heater metal line.

The measurement of the Seebeck coefficient follows a similar procedure (as discussed in the main text); a current passed through the heater electrode induces a temperature gradient in the diaphragm, resulting in a Seebeck voltage across the gated 2DEG mesa that spans the suspension region (Figure 2c). Using a similar calibration procedure for the heater line, the temperature drop across the mesa can be used to extract the Seebeck coefficient. As we shall demonstrate, the measured Seebeck coefficient corresponds to the 2DEG contribution exclusively since the III-V buffer layers are semi-insulating and the temperature drop in the silicon supported region (and hence the silicon contribution) can be neglected.

**Temperature-Resistance Calibration Procedure:**

Figure S4a shows the test setup used to measure the in-plane thermal conductivity of the AlGaN/GaN hetero-structure. In order to ensure accuracy in the thermal conductivity measurements, we performed careful resistance versus temperature calibration for the Ti/Pt heater and sensor lines. For the heater line, a DC current source (Keithley 2400), with a current value of ~100 µA and a DC voltage source (Agilent 34401) were used to measure the resistance of the Ti/Pt trace. To measure the resistance of the sensor line, we used an AC voltage lock-in amplifier (Zurich Instruments HF2LI) with a lock-in frequency of 97 Hz. AC voltage measurement across a fixed resistor (1 kΩ, ultra-low TCR of less than 1 ppm) was used to infer the AC current from the applied AC voltage. The lock in-amplifier was chosen for the sensor side to minimize self-heating effects and block environmental noise. In order to calibrate the resistances of both lines, the substrate of the suspended membrane was attached to a temperature controlled chuck using high vacuum thermal grease (Apiezon, Inc.). A current amplitude of ~100 µA was carefully chosen for the purpose of resistance calibration to avoid self-heating effects in the sensor line.

Figure S4b shows the calibration curves of resistance of the heater and sensor lines from 50 K to 300 K. This calibration curve is later used to extract the temperature of the heater line (when heating power is applied to it) and sensor line to extract the thermal conductivity of the AlGaN/GaN heterostructure. Note that the plotted resistance values are obtained by averaging over 20 measurements spaced by 2 seconds at each substrate temperature. In each case, the error bar (defined as the range) for the resistance measurement is smaller than the size of the markers. A similar calibration procedure was performed for the heater line in the Seebeck coefficient measurement platform for the thick and thin GaN samples, as can be seen in Figures S4c and S4d.

**Thermal Conductivity Extraction Procedure:**

We first focus on the thermal conductivity extraction procedure. In our device, the heater and sensor lines have a width ($W$) of 5 µm, and are spaced by a distance ($D_{HS}$) of 75 µm (center to center), as seen in Figure S3a. As highlighted previously, the location of the heater and sensor resistances on the suspended membrane ($R_H$ and $R_S$) are chosen such that the heat transfer can be well approximated as 1-D. Figure S5a shows a cross-section schematic of the thermal resistance network with the different pathways for heat sinking when a current is applied to the heater metal. Since we have established that thermal conduction is the only heat transport mechanism that needs to be accounted for, the thermal resistance of the suspended film ($R_F$) can be written as:

$$R_F = \frac{2(T_H - T_S)}{P_H} - 2R_{ox} - \frac{2(R_{mox} + R_{oxg})}{A_H} \tag{S1}$$

where $T_H$ and $T_S$ are the heater and sensor line temperatures, $P_H$ is the input power to the heater and $R_{ox}$ is the thermal resistance of the Al$_2$O$_3$ layer, $A_H$ is the area projected area of the heater electrode (5 μm × 200 μm), $R_{mox}$ is the thermal boundary resistance of the heater/Al$_2$O$_3$ interface and $R_{oxg}$ is the thermal boundary resistance of the Al$_2$O$_3$/GaN interface. The thermal conductivity of the film can be extracted from $R_F$ and the known film dimensions. To calculate the thermal resistances, we denote $t_{ox}$ and $t_F$ as the thicknesses of the alumina and AlGaN/GaN/buffer film, respectively. We used a thermal boundary resistance of 2.8×10$^{-8}$ m$^2$K W$^{-1}$ for $R_{mox}$.[7] Although an experimental determination of the thermal boundary resistance across the Al$_2$O$_3$/GaN film interface is not available, we estimated $R_{oxg} \approx$ 1×10$^{-8}$ m$^2$KW$^{-1}$ based on measurements across amorphous dielectric/Si interfaces,[8] since GaN and Si have similar Debye temperatures.[9] The thermal resistance of the alumina layer can be estimated as $R_{ox} = t_{ox}/(k_{ox}A_H)$, where $k_{ox}$ is the temperature dependent thermal conductivity of the alumina layer. The measurements for the thermal conductivity of amorphous alumina films have been published in the literature before. It is worth noting that amorphous films are typically modeled by the differential effective-medium (DEM) approximation, where $k \propto n^{\frac{2}{3}}$, with $n$ denoting the atomic density of the film.[10] Thus, the variation in the thermal conductivities between the different films may be associated with different densities, which depends strongly on the growth technique and deposition temperature. Our film is prepared via atomic layer deposition (ALD) at a temperature of 200°C. Thermal conductivity of films made by this process has been previously measured by Gorham *et al.* at room temperature.[10] The temperature dependent thermal conductivities of alumina films prepared under different conditions have been reported by a few other research groups,[7,10–12] as seen in Figure S3b. Attributing the difference exclusively to density variations, we fit the thermal conductivity obtained by Lee *et al.* for different temperatures,[12] and scale it to match the value obtained by Gorham *et al.* at room temperature[10] to obtain $k_{ox}$, marked by a red line in Figure S5b. In conclusion, since $R_{mox}$, $R_{oxg}$ and $R_{ox}$ are known from Equation S1, and the thickness of the heterostructure was determined from SEM measurements of the cross-section, we can calculate $R_F$ and thus obtain the thermal conductivity ($k_F$) of the suspended film.

Figure S6 shows the typical thermal conductivity measurement procedure for our films. These plots are from experiments with the thick GaN sample. In this experiment, the substrate is held at 300 K. The sensor is maintained at the calibration current of ~100 μA (Figure S6f), while the heater current is ramped up in a half-sinusoid from its initial calibration value (Figure S6b). Before each resistance measurement, we wait for 2 seconds after the current ramp to allow the system to equilibrate. The waiting interval of 2 seconds was chosen based on an estimation of a thermal time constant of ~2 milli-seconds for the suspended membrane from COMSOL simulations. The heater & sensor temperature (converted from the resistance via the calibration curve in Figure S6b) track the current pattern, with the initial temperature equal to the substrate temperature, as seen in Figure S6c and Figure S6d. The extracted temperature difference can be used to calculate the in-plane film thermal conductivity via Equation S1, after accounting for the $Al_2O_3$ temperature drop, as discussed previously. At each substrate temperature, currents from 75% of the peak current value to the peak current value (7.5 mA to 10 mA in Figure S6b) are used to obtain the thermal conductivity (at that substrate temperature), which results in the error bars shown in Figure 2e of the main text. We also note that hysteresis did not occur in our heater and sensor lines. This can be seen from the temperature versus power lines in Figure S6a and Figure S6e, which overlap in the temperature ramp and cool cycles.

**Seebeck Coefficient Extraction Procedure:**

Figure S7 shows a typical Seebeck coefficient measurement procedure. Similar to the thermal conductivity measurement, the heater current is ramped up from its calibration value, setting up a lateral temperature gradient along the 2DEG mesa which translates to a measurable Seebeck voltage (Figure S7d). At each substrate temperature, currents from 75% of the peak current value to the peak current value (7.5 mA to 10 mA in Figure S7c) are used to obtain the Seebeck coefficient (at that substrate temperature), resulting a small error bar in the measured values. The Seebeck coefficient of the 2DEG is given as $S = V_{2DEG}/(T_1 - T_2)$, as depicted in Figure S7d and Figure S7e. $T_1$ is the temperature extracted 30 μm away from the center of the heater line, where the 2DEG mesa begins. $T_1$ is related to the heater temperature $T_H$ as:

$$\frac{(T_H - T_1)}{P_H} = R_{ox} + \frac{R_F}{2} + \frac{(R_{mox} + R_{oxg})}{A_H} \quad \text{(S2)}$$

where $R_F$ is calculated using the measured film thermal conductivity and a length of 30 μm ($D_S$, depicted in Figure S3d) and $R_{ox}$ is calculated as discussed earlier. $T_2$ is the substrate temperature. Knowing $T_1, T_2$ and $V_{2DEG}$, the total Seebeck coefficient $S$ can be extracted.

The temperature at the contact outside the suspended region ($T_2$) is assumed to be at the substrate temperature.

The temperature drop in the silicon supported region is a small fraction of the total temperature drop, thus, the contribution to the Seebeck coefficient from the supported region can be ignored. Most importantly, this ensures that any contribution from the silicon in the supported region to the measured Seebeck coefficient can be ignored. We can estimate the temperature drop in the silicon supported region via simple finite-element simulations (performed in COMSOL), since we know the thermal properties of the suspended heterostructure and $Al_2O_3$ layers.

Figure S8b shows the simulated temperature profile from the center of the suspension region to the silicon supported region for the thick GaN sample, assuming that the base of the silicon is held at 300 K. In this simulation, we use the determined thermal conductivity of the heterostructure (110 $Wm^{-1}K^{-1}$ at 300 K) and the $Al_2O_3$ layer (1.7 $Wm^{-1}K^{-1}$). The simulated current in the Pt heater is 16 mA. The room temperature thermal conductivity of the silicon is assumed to be 156 $Wm^{-1}K^{-1}$.[2] It can be seen that a small fraction, ~2.7% of the total temperature drop is across the silicon supported region. A similar simulation using the properties of the thin GaN sample (58 $Wm^{-1}K^{-1}$ at 300 K) and a heater current of 9 mA shows that only ~1.1% of the total temperature drop is across the silicon supported region (Figure S8c). At measurement temperatures below 300 K, the fraction of temperature dropped across the silicon supported region will be much smaller than the room temperature values, because of the large increase in the thermal conductivity of silicon,[2] as it is not limited by boundary scattering unlike the heterostructure layers. This can be seen in Figure S8d, where we compare the temperature dependent thermal conductivities of the thin and thick GaN sample against thermal conductivity data for p-type bulk silicon doped at $4 \times 10^{16}$ $cm^{-3}$.[2] These thermal conductivity values are appropriate for our Si substrate, which is 725 μm thick, p-type and doped between $10^{16}$-$10^{17}$ $cm^{-3}$.

**Supplementary Note 3:** Mobility Model

To model the mobility in the AlGaN/GaN 2D electron gas, we need to understand the scattering rates for the electrons in the 2DEG quantum well with phonons (acoustic and optical), and with roughness of the 2DEG interface. The electronic state for a 2D quantum well can be described by wave vector $\boldsymbol{k} = (k_x, k_y)$ in the plane of the AlGaN/GaN interface, and subband index $n$ to describe the wave function along the confinement direction ($z$). Under this assumption, we can write the wave function and electron energy for the electrons in the bottom subband as:

$$\Psi_{n,k} = \psi(z)e^{ik\cdot r} \tag{S3}$$

$$E_n(k) = E_n + \frac{\hbar^2 k^2}{2m^*} \tag{S4}$$

where $r = (x, y)$ denotes the spatial coordinate in-the-plane of the 2DEG and $E_n$ is the energy at the bottom of the subband corresponding to index $n$.[13] Figure S9a shows $E_n$ as a function of temperature (50 to 300 K) with respect to the Fermi level for the bottom two subbands for the thin and thick GaN sample. Since the majority of conduction electrons (> 90 %) are in the lower subband (estimated from the subband energies in Figure S9a), we only consider the bottom subband ($n = 1$) for evaluating all the electronic transport properties (thus neglecting inter-subband scattering). To model the wave function along the confinement direction in Equation S3, we can use the Fang-Howard expression, where $\psi(z) = \sqrt{\frac{b^3 z^2}{2}} e^{-\frac{bz}{2}}$.[3] Here, the parameter $b = \left(\frac{12 m^* e^2 n_{\text{eff}}}{\epsilon \hbar^2}\right)^{\frac{1}{3}}$, where $n_{\text{eff}} \approx \frac{11}{32} n_{2\text{D}}$,[3] assuming that the barrier layer is un-doped and all the 2DEG electrons are a result of built-in polarization fields at the AlGaN/GaN interface.

The scattering rates for electrons can be evaluated using Fermi's golden rule, for which we need to calculate the matrix elements with the correct scattering potentials for the different mechanisms. For the sake of brevity, we will skip the details, which can be found elsewhere.[3] In our scattering picture, the 3D phonon can be represented by the wave vector $Q = (q, q_z)$, where $q$ and $q_z$ represent the in-plane and out-of-plane component. When an electron with initial wave vector $k = (k_x, k_y)$ scatters with a phonon with wave vector $Q$, its final state can be written as $k' = k + q$ from conservation of momentum in-plane. If the collision is elastic, we can write $|q| = q = 2k \sin\left(\frac{\theta}{2}\right)$, where $\theta$ is the angle between $k$ and $k'$. The in-plane scattering matrix elements ($M$) are identical to the ones used in for scattering with 3D electrons. However, in this case, because we need to account for the out-of-plane phonon wave vector $q_z$, the 2D matrix scattering element is modified as

$$M_{2D}^2 = \int M^2 |I(q_z)|^2 \, dq_z, \tag{S5}$$

where $I(q_z) = \int \psi(z)^2 e^{iq_z z}$. Using the Fang-Howard form for $\psi(z)$, $|I(q_z)|^2$ can be simplified as $\frac{b^6}{(b^2 + q_z^2)^3}$.[3] For the purposes of calculating the AlGaN/GaN mobility, the mechanisms we consider here are scattering from acoustic phonons, optical phonons and roughness at the AlGaN/GaN quantum well interface. In particular, scattering by ionized impurities is neglected since the AlGaN barrier layer is assumed to be un-doped. Further, only acoustic scattering via the deformation potential is considered and piezoelectric scattering is neglected as it has previously been found to be negligible for the purpose of evaluating the mobility.[3]

Screening of the electron-phonon interaction by the carriers themselves is important to consider for the elastic processes (in our case, for acoustic phonon scattering and roughness scattering). This is often done by scaling the matrix scattering element $M_{2D}$ by the screening function, defined as[3]

$$S(q,T) = 1 + \frac{e^2 F(q) \Pi(q,T)}{2\epsilon q}, \tag{S6}$$

where $q = |\mathbf{q}|$, $F(q)$ and $\Pi(q,T)$ are the form factor and the polarizability function whose definitions are well known in the literature.[13] Once $S(q)$ is known, we can calculate the scattering times $\tau(E)$ for the 2DEG electrons as functions of electron kinetic energy ($E$). The integrated expressions for $\tau(E)$ over the limits of the scattering angle $\theta$ (from 0 to $2\pi$) for acoustic deformation potential scattering, $\tau_{ac}(E)$ and optical phonon scattering, $\tau_{opt}(E)$ can be found in former work.[3] For roughness scattering, we correct the expression found in former work[3] (missing a factor of $\pi$), to get

$$\frac{1}{\tau_{ir}(E)} = \frac{m^* \Delta^2 \lambda^2 e^4 (n_{2D})^2}{8\pi \hbar^3 \epsilon^2} \int_0^{2\pi} e^{\frac{-q^2 \lambda^2}{4}} \frac{(1 - \cos\theta)}{S(q,T)^2} d\theta, \tag{S7}$$

where $\Delta$ is the RMS roughness of the interface and $\lambda$ is a parameter defined as the auto-correlation length.[3] In order to accurately fit the mobility data over temperature, we set $\lambda = 7.5$ nm, and values of $\Delta$ corresponding to 1 nm and 2 nm for the thick GaN and thin GaN sample, respectively. An AFM image of the sample surface is shown in Figure S9b, where the RMS roughness is found to be in this range (~1.4 nm). Once the values for the different scattering times are obtained, the total scattering time $\tau(E)$ can be calculated by adding up the different scattering rates. Finally, we calculate the energy averaged scattering time as a function of temperature as

$$\tau_{av}(T) = \frac{\int \tau(E) \frac{\partial f_0(E)}{\partial E} dE}{\int E \frac{\partial f_0(E)}{\partial E} dE}, \tag{S8}$$

where $f_0(E)$ is the Fermi function and the limits of integration are from the subband bottom $E_1$ to $\infty$ (referenced to $E_F$). Since $n_{2D} \approx \frac{m^*(E_F - E_1)}{\pi \hbar^2}$ when using only the bottom subband, we obtain $E_1 \approx$ -108 meV, which is consistent with the Schrödinger–Poisson model (Figure S9a).

Once $\tau_{av}(T)$ is calculated from Equation S8, the 2DEG mobility for both the experimental samples can be obtained.

**Supplementary Note 4:** Phonon Drag Model

As discussed in main paper, the expression for phonon drag for the case of 3D phonons interacting with 2D electrons is

$$S_{\text{ph}} = -\frac{(2m^*)^{\frac{3}{2}} v_{av}^2}{4(2\pi)^3 k_B T^2 n_{2D} e \rho} \int_0^\infty dq \int_{-\infty}^\infty dq_z \frac{\Xi^2(\mathbf{Q}) q^2 Q^2 |I(q_z)|^2 G(\mathbf{Q}) \tau_{\text{ph}}(\mathbf{Q})}{S^2(q,T) \sinh^2\left(\frac{\hbar \omega_Q}{2 k_B T}\right)}. \quad (S9)$$

The definitions for $S(q,T)$ and $I(q_z)$ follow from Supplementary Note 3. The explicit expression for $\Xi(\mathbf{Q})$ is[14]

$$|\Xi|^2 = D^2 + \frac{8 q_z^2 q^2 + q^4}{2(q^2 + q_z^2)^2}, \quad (S10)$$

where the first term represents the scattering via the deformation potential (with strength of the interaction described by constant $D$) and the second term accounts for piezoelectric scattering. In Equation S9, $G(\mathbf{Q})$ is the energy integral, which is written as:

$$G(\mathbf{Q}) = \frac{1 - e^{\frac{-\hbar \omega_Q}{k_B T}}}{\hbar \omega_Q} \times \int_\gamma^\infty dE \frac{f_0(E)(1 - f_0(E + \hbar \omega_Q))}{\sqrt{E - \gamma}}. \quad (S11)$$

In Equation, S11, $\gamma = \frac{(\hbar \omega_Q - E_q^2)}{4 E_q}$, with $E_q = \frac{\hbar^2 q^2}{2m^*}$. In Figure S9c, we show the form $I(q_z)$ for a range for $n_{2D}$ varying from $10^{12}$ to $10^{13}$ cm$^{-2}$. For small values of $q_z$, $I(q_z) \approx 1$, but around $q_z$ corresponding to the Debye wavelength (about $1.55 \times 10^{10}$ m$^{-1}$ in GaN), $I(q_z) \approx 0$. The physical interpretation is that for thinner quantum wells (smaller $\Delta z$), larger values of $q_z$ are allowed to interact with the 2D electrons because the momentum conservation in the out-of-plane direction is less stringent.[15] Finally, because of the specific shape of $I(q_z)$, we can set the limits of the integration in Equation S9 to the Debye wave vector (instead of $\infty$).

**Supplementary Note 5:** Thermal Conductivity Model

As seen in Supplementary Note 4, to accurately calculate $S_{\text{ph}}$ via Equation S9, it is necessary to calculate the phonon scattering time, $\tau_{\text{ph}}(\mathbf{Q})$ in the GaN layer. This can be estimated accurately from the in-plane thermal conductivity measurements of the suspended AlGaN/GaN film. Since we do not have thermal conductivity measurements of the GaN layer exclusively, we follow a more involved approach. In particular, we first model the thermal conductivity of the composite film and compare with experimental data. Then, we use the model for the GaN film to estimate the $\tau_{\text{ph}}(\mathbf{Q})$ needed to calculate $S_{\text{ph}}$.

Since our suspended film is a composite consisting of an AlN layer, $Al_xGa_{1-x}N$ transition layers and a GaN layer, the overall thermal conductivity ($k$) can be estimated as $\sum k_i t_i / \sum t_i$, where $k_i$ and $t_i$ refer to the thermal conductivities and thicknesses of individual layers. For each multilayer, we use a Boltzmann Transport Equation (BTE) model to quantify $k_i$ with layer thickness ($t_i$). Using a simple Debye approximation for the phonon dispersion with an average velocity over the acoustic phonon modes ($v_{\text{av}}$), the in-plane thermal conductivity for each layer can be written as[16]

$$k_i = \frac{k_B^4 T^3}{2\pi^2 \hbar^3 v_{\text{av}}} \int_0^{\theta_D/T} \frac{x^4 e^x \tau(x)}{(e^x - 1)^2}, \quad (S12)$$

where $\theta_D$ is the Debye temperature for the multilayer, $T$ is the temperature, and $x = \hbar\omega/(k_B T)$. Here, $\omega$ is the phonon frequency, which can be approximated as $v_{\text{av}}\sqrt{q^2 + q_z^2}$ assuming a 3-D isotropic linear phonon dispersion. The total phonon scattering time $\tau$ is calculated by Mathiessen's rule with contributions from Umklapp ($\tau_U$), alloy ($\tau_A$), boundary ($\tau_B$) and defect scattering ($\tau_D$), respectively. Phonon-phonon scattering is evaluated using via the relaxation term for Umklapp processes[17]

$$\tau_U(x) = \frac{P k_B^2 T^3 x^2 e^{-\frac{C_U}{T}}}{\hbar^2}, \quad (S13)$$

where the constants $P$ and $C_U$ for GaN and AlN are listed in Table S1. Scattering with impurities is neglected since its effect is found to be negligible for unintentionally doped films.[18] For the $Al_xGa_{1-x}N$ layers, all the material parameters (e.g. $v_{\text{ac}}$, $\theta_D$, $P$, $C_U$) are averaged over the AlN and GaN fractions, in context of the virtual crystal model.[4] Alloy scattering severely reduces the thermal conductivity of the transition layers and is evaluated as a point defect scattering term.[19] For the sake of brevity, we skip the details, which can be found in Liu *et al.*[19] The defect scattering term ($\tau_D$) included core, screw, edge and mixed dislocations with total density $N_{\text{dis}}$, whose effect is to reduce the thermal conductivity.[20]

Although we have a composite film (and thus, the dislocation density is expected to vary for the different layers), we estimated an average dislocation density for the suspended film via X-Ray Diffraction (XRD) measurements. These values were estimated to be $\approx 9 \times 10^8$ cm$^{-2}$ and $\approx 2.5 \times 10^9$ cm$^{-2}$ for the thick and thin GaN samples, the details of which can be found in former work.[1] To evaluate the boundary scattering term, we used $\tau_B \approx 2.38 t_i/v_{\text{av}}$, which is a model that is used for nanowires,[18] but will suffice to model the dependencies observed in the measured thermal conductivity with temperature.

**Supplementary Note 6:** Simulation Codes

The codes to simulate the diffusive Seebeck coefficient, the phonon drag component of the Seebeck coefficient and the thermal conductivity are available (as MATLAB files) at: https://github.com/ananthy/PhononDrag.